\PassOptionsToPackage{unicode}{hyperref}
\PassOptionsToPackage{hyphens}{url}
\PassOptionsToPackage{dvipsnames,svgnames,x11names}{xcolor}
\documentclass[
]{article}
\usepackage{amsmath,amssymb}
\usepackage{lmodern}
\usepackage{iftex}
\ifPDFTeX
  \usepackage[T1]{fontenc}
  \usepackage[utf8]{inputenc}
  \usepackage{textcomp} 
\else 
  \usepackage{unicode-math}
  \defaultfontfeatures{Scale=MatchLowercase}
  \defaultfontfeatures[\rmfamily]{Ligatures=TeX,Scale=1}
\fi
\IfFileExists{upquote.sty}{\usepackage{upquote}}{}
\IfFileExists{microtype.sty}{
  \usepackage[]{microtype}
  \UseMicrotypeSet[protrusion]{basicmath} 
}{}
\makeatletter
\@ifundefined{KOMAClassName}{
  \IfFileExists{parskip.sty}{%
    \usepackage{parskip}
  }{
    \setlength{\parindent}{0pt}
    \setlength{\parskip}{6pt plus 2pt minus 1pt}}
}{
  \KOMAoptions{parskip=half}}
\makeatother
\usepackage{xcolor}
\usepackage{graphicx}
\makeatletter
\def\maxwidth{\ifdim\Gin@nat@width>\linewidth\linewidth\else\Gin@nat@width\fi}
\def\maxheight{\ifdim\Gin@nat@height>\textheight\textheight\else\Gin@nat@height\fi}
\makeatother
\setkeys{Gin}{width=\maxwidth,height=\maxheight,keepaspectratio}
\makeatletter
\def\fps@figure{htbp}
\makeatother
\providecommand{\tightlist}{%
  \setlength{\itemsep}{0pt}\setlength{\parskip}{0pt}}
\newlength{\cslhangindent}
\newlength{\csllabelwidth}
\newlength{\cslentryspacingunit} 
\newenvironment{CSLReferences}[2] 
 {
  \setlength{\parindent}{0pt}
  \ifodd #1
  \let\oldpar\par
  \def\par{\hangindent=\cslhangindent\oldpar}
  \fi
  \setlength{\parskip}{#2\cslentryspacingunit}
 }%
 {}
\usepackage{calc}

\ifLuaTeX
\usepackage[bidi=basic]{babel}
\else
\usepackage[bidi=default]{babel}
\fi
\babelprovide[main,import]{american}

\def\languageshorthands#1{}
\ifLuaTeX
  \usepackage{selnolig}  
\fi
\IfFileExists{bookmark.sty}{\usepackage{bookmark}}{\usepackage{hyperref}}
\IfFileExists{xurl.sty}{\usepackage{xurl}}{} 
\hypersetup{
  pdftitle={NOMAD CAMELS: Configurable Application for Measurements,
Experiments and Laboratory Systems},
  pdfauthor={Alexander D. Fuchs, Johannes A. F. Lehmeyer, Heinz Junkes,
Heiko B. Weber, Michael Krieger},
  pdflang={en-US},
  colorlinks=true,
  linkcolor={Maroon},
  filecolor={Maroon},
  citecolor={Blue},
  urlcolor={Blue},
  pdfcreator={LaTeX via pandoc}}

\title{NOMAD CAMELS: Configurable Application for Measurements,
Experiments and Laboratory Systems}

\usepackage[affil-it]{authblk}
\usepackage{orcidlink}
\author[1,2%
  *%
  ]{Alexander D. Fuchs%
    \,\orcidlink{0000-0003-1896-9242}\,%
    }
\author[1,2%
  *%
  ]{Johannes A. F. Lehmeyer%
    \,\orcidlink{0000-0003-2041-9987}\,%
    }
\author[3%
  ]{Heinz Junkes%
    \,\orcidlink{0000-0002-0218-4873}\,%
    }
\author[1%
  ]{Heiko B. Weber%
    \,\orcidlink{0000-0002-6403-9022}\,%
    }
\author[1%
  \ensuremath\mathparagraph]{Michael Krieger%
    \,\orcidlink{0000-0003-1480-9161}\,%
    }

\affil[1]{Lehrstuhl für Angewandte Physik, Department Physik,
Friedrich-Alexander Universität Erlangen-Nürnberg, Germany.}
\affil[2]{Physics Department and CSMB, Humboldt-Universität zu Berlin,
Berlin, Germany.}
\affil[3]{Fritz-Haber-Institut, Berlin, Germany.}
\affil[$\mathparagraph$]{Corresponding author}
\affil[*]{These authors contributed equally.}
\date{20 November 2023}

\begin{document}
\maketitle

\hypertarget{summary}{%
\section{Summary}\label{summary}}

NOMAD CAMELS (short: CAMELS) is a configurable, open-source measurement
software that records fully self-describing experimental data. It has
its origins in the field of experimental physics where a wide variety of
measurement instruments are used in frequently changing experimental
setups and measurement protocols. CAMELS provides a graphical user
interface (GUI) which allows the user to configure experiments without
the need of programming skills or deep understanding of instrument
communication. CAMELS translates user-defined measurement protocols into
stand-alone executable Python code for full transparency of the actual
measurement sequences. Existing large-scale, distributed control systems
using e.g.~EPICS can be natively implemented. CAMELS is designed with
focus on full recording of data and metadata. When shared with others,
data produced with CAMELS allow full understanding of the measurement
and the resulting data in accordance with the \textbf{FAIR}
(\textbf{F}indable, \textbf{A}ccessible, \textbf{I}nteroperable and
\textbf{R}e-usable) principles
(\protect\hyperlink{ref-Wilkinson2016}{Wilkinson et al., 2016}).

\hypertarget{statement-of-need}{%
\section{Statement of need}\label{statement-of-need}}

Research data management has piqued greater and greater interest in
recent years. Today, research funding agencies demand sustainable
research data strategies. The key criterion is to create research data
following the FAIR principles and thereby improve world-wide data-driven
research (\protect\hyperlink{ref-DFG_position_paper}{\emph{DFG Position
Paper}, 2018}). While one ingredient, electronic lab notebooks, are an
important step towards FAIR data, it is equally important to record the
measurement data along FAIR principles as early as possible in the
research workflow.

In experimental physics many custom-built measurement setups are
controlled by very specific software written by individual researchers.
This results in a heterogeneous landscape of software fragments for
measurements written in many different languages and with often
incomplete documentation, making it almost impossible for other
researchers to extend existing code. The degree to which the stored raw
data is understandable varies greatly but is often unintelligible even
for researchers from the same lab. Important metadata such as instrument
settings or the actual measurement steps performed to obtain the final
raw data are rarely recorded, making it virtually impossible to
reproduce experiments. Therefore, the documentation of experiments is
incomplete preventing FAIR research data. Although there are some tools
available (e.g.~\emph{SweepMe!}
(\protect\hyperlink{ref-SweepMe}{\emph{SweepMe}, n.d.}), \emph{iC}
(\protect\hyperlink{ref-pernstich2012}{Pernstich, 2012}), PyMoDAQ
(\protect\hyperlink{ref-PyMoDAQ}{\emph{PyMoDAQ}, n.d.})) to realise
control of arbitrary measurement instruments, they are frequently not
open-source or their data output is not compliant to the FAIR
principles.

\begin{figure}
\centering
\includegraphics[width=0.8\textwidth,height=\textheight]{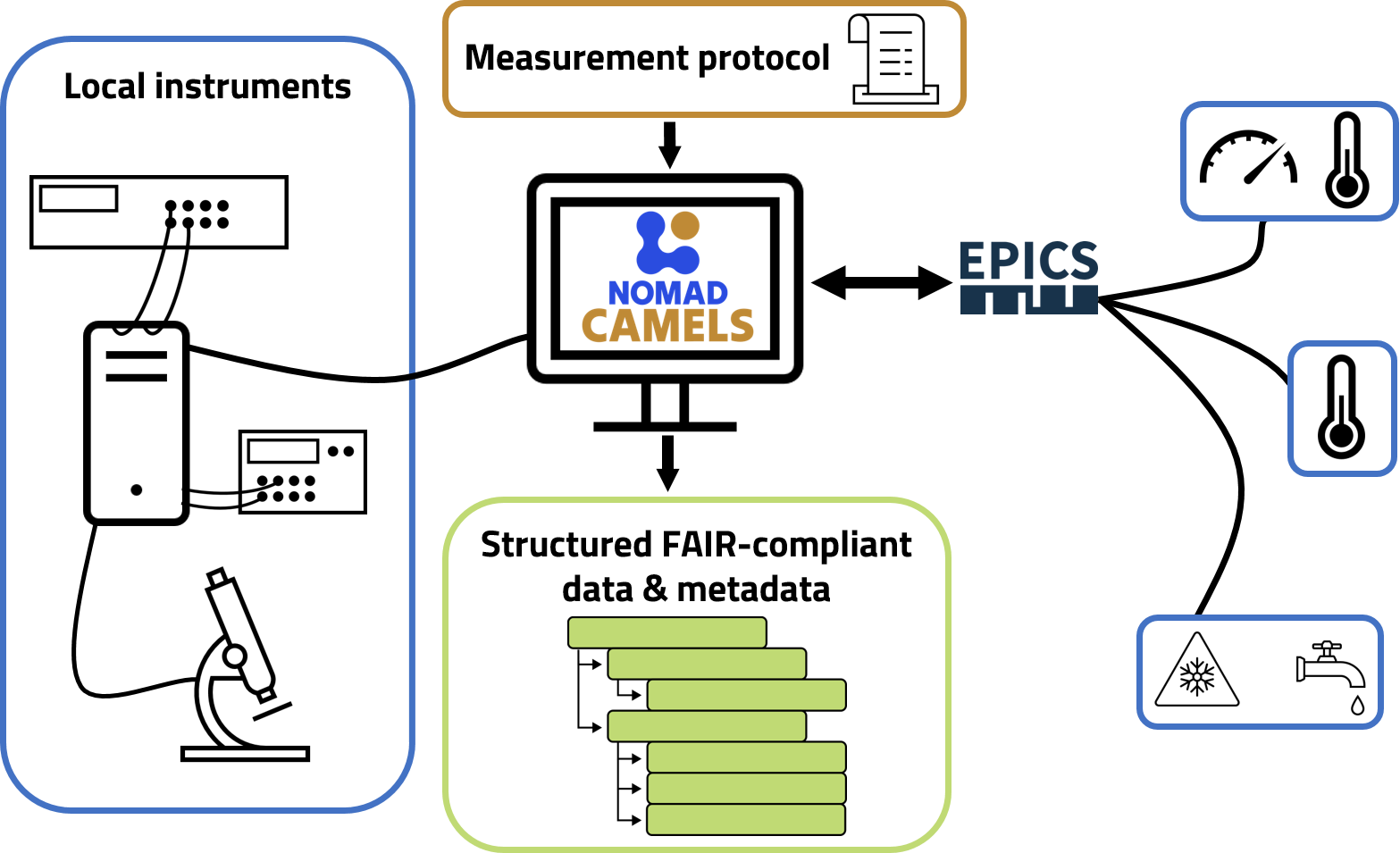}
\caption{Visualization of CAMELS functionality and workflow. CAMELS
connects directly with local instruments and/or large-scale lab
infrastructure running network protocols, e.g.~EPICS. Customizable
measurements protocols are translated into Python code and executed. The
output is FAIR-compliant measurement data. \label{fig:camels_overview}}
\end{figure}

\hypertarget{nomad-camels}{%
\section{NOMAD CAMELS}\label{nomad-camels}}

CAMELS is an open-source tool that automatically collects all
computer-accessible experimental metadata. It features a user-friendly
graphical interface that enables the creation and customization of
measurements without the need for programming knowledge. By default, the
data is stored in a structured HDF5 file format that closely resembles
the structure of the NeXus standard
(\protect\hyperlink{ref-FAIRmatNeXusProposal}{\emph{FAIRmat NeXus
Proposal}, n.d.}; \protect\hyperlink{ref-Konnecke2015}{Könnecke et al.,
2015}). The final HDF5 file contains both the actual measurement data
and metadata in a single file, compliant to the FAIR principles.

Moreover, CAMELS allows for direct access to the \emph{NOMAD} repository
(\protect\hyperlink{ref-scheidgenFAIRResearchData2023}{Scheidgen,
Brückner, et al., 2023}),
(\protect\hyperlink{ref-scheidgenNOMADDistributedWebbased2023}{Scheidgen,
Himanen, et al., 2023}) or its on-premise installation called
\emph{NOMAD Oasis} enabling direct linking to electronic lab notebook
(ELN) entries. The user can for example connect measurements to previous
experiment workflows documented in \emph{NOMAD} ELNs. CAMELS can
subsequently upload measurement results directly into the ELN providing
a simple and stream-lined data workflow.

CAMELS builds on \emph{bluesky} (\protect\hyperlink{ref-Allan2019}{Allan
et al., 2019}; \protect\hyperlink{ref-bluesky}{\emph{Bluesky}, n.d.})
initially developed to control instruments at large-scale research
institutions using EPICS (\protect\hyperlink{ref-EPICS}{\emph{EPICS},
n.d.}; \protect\hyperlink{ref-Knott1994}{Knott et al., 1994}). In
CAMELS, \emph{bluesky} is employed to orchestrate the instrument
communication either directly (e.g.~via PyVISA
(\protect\hyperlink{ref-PyVISA}{\emph{PyVISA}, n.d.})) or via using
network protocols. Existing lab infrastructure controlled by EPICS is
therefore immediately accessible. A schematic overview of the
functionality of CAMELS is displayed in \autoref{fig:camels_overview}.

CAMELS provides a comprehensive set of functionalities that can be split
into three primary components: instrument management, measurement
protocols and manual controls.

\begin{figure}
\centering
\includegraphics{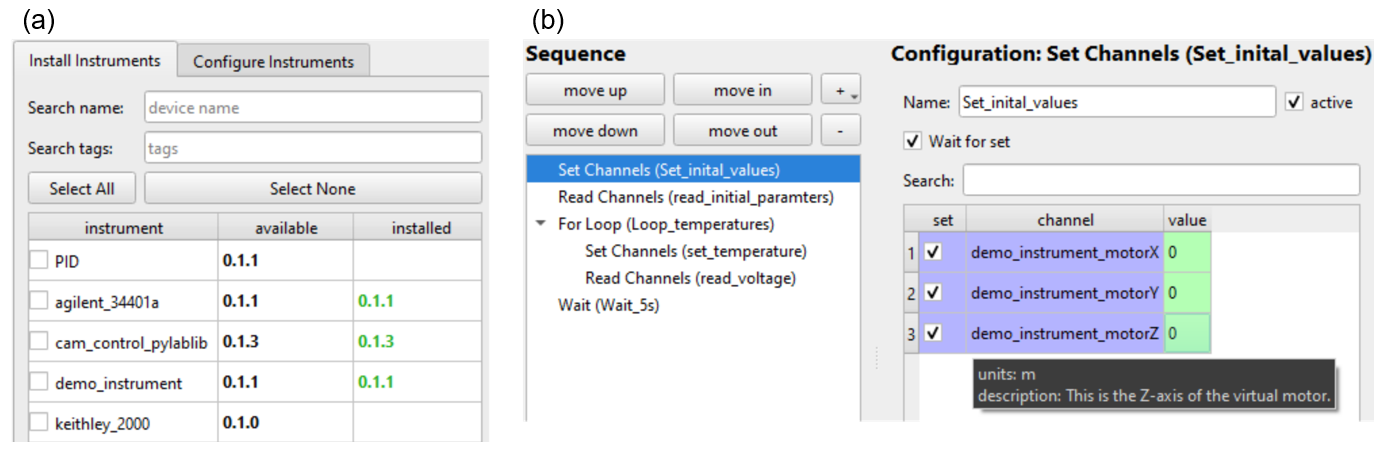}
\caption{\textbf{(a)} The instrument manager allows to install and
configure instrument drivers from the curated instrument driver library
or the user's hard drive. \textbf{(b)} The measurement protocol editor
allows users to configure arbitrary measurement sequences.
\label{fig:manager_protocols}}
\end{figure}

\hypertarget{instrument-management}{%
\subsection{Instrument Management}\label{instrument-management}}

Scientific instruments can be added to CAMELS in two ways: The first
involves the \emph{instrument manager} (c.f.
\autoref{fig:manager_protocols}a) to add instruments from the official
curated driver repository
(\protect\hyperlink{ref-CAMELS_drivers}{{``CAMELS-Drivers,''} n.d.}).
These drivers are installed into the Python environment via \emph{pip}
(\protect\hyperlink{ref-PipDocumentationV23}{\emph{Pip}, n.d.}) with
each driver being packaged individually.

The second way is to add self-built drivers by creating the necessary
files locally and placing these in the directory specified in the CAMELS
settings. To facilitate this process CAMELS provides a \emph{driver
builder} that automatically generates the essential structure and
boilerplate code. As CAMELS is an open-source project developed by and
for the community, users are encouraged to contribute to the driver
library by creating pull requests for new drivers on the GitHub
repository (\protect\hyperlink{ref-CAMELS_drivers}{{``CAMELS-Drivers,''}
n.d.}).

In general, a CAMELS driver comprises two files: One containing the
hardware interface communication, the other one defining the available
instrument settings. Data communication to instruments is handled via
\emph{channels} that can be set and/or read; they correspond to an
instrument's individual functionality or physical property.

\hypertarget{measurement-protocols}{%
\subsection{Measurement protocols}\label{measurement-protocols}}

In CAMELS a \emph{measurement protocol} is a distinctive sequence of
individual steps including setting and reading instrument channels (see
\autoref{fig:manager_protocols}b), loops, conditional execution, running
sub-protocols, PID control, etc. This yields a measurement in a
`recipe-style' format, where the next step is usually executed after the
successful completion of the preceding step. Asynchronous data
acquisition is supported.

CAMELS translates the protocol created in the GUI into a Python script,
which is then executed. The script can be viewed, run independently of
CAMELS, and modified if required. CAMELS protocols and settings can be
stored and shared with colleagues enabling easy repeatability of
experiments.

\hypertarget{manual-controls}{%
\subsection{Manual controls}\label{manual-controls}}

Certain scientific instruments require manual control before starting
predefined measurement routines, e.g.~adjusting stages, controlling
temperature, valves, pumps, etc. In CAMELS this is achieved through the
\emph{manual controls} functionality which can be applied to any
writable instrument channel.

\hypertarget{data-output}{%
\subsection{Data output}\label{data-output}}

\begin{figure}
\centering
\includegraphics{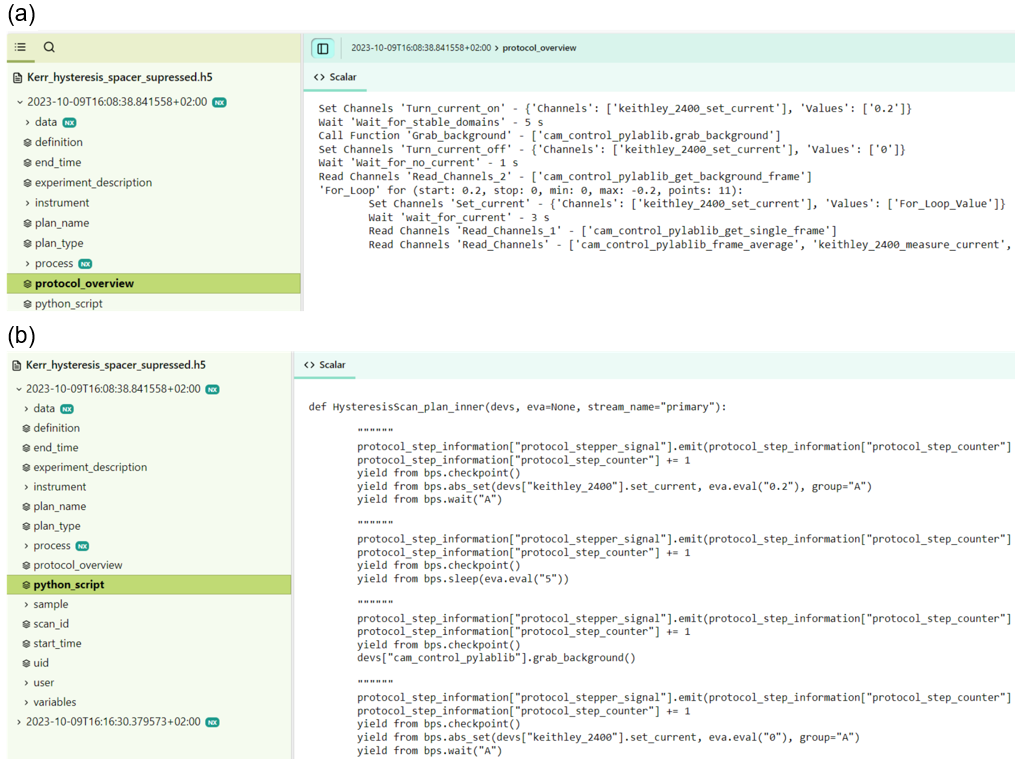}
\caption{CAMELS stores measurement data together with rich metadata
collected automatically into a structured HDF5 file by default. This
includes \textbf{(a)} a human readable measurement protocol summary and
\textbf{(b)} the executable Python script that was used to actually
record the data. This allows others to understand the data acquisition
and to reproduce the experiment. \label{fig:h5_data}}
\end{figure}

After executing the measurement protocol, the time-stamped data is by
default saved to an HDF5 file with a structure similar to the NeXus
standard (\protect\hyperlink{ref-Konnecke2015}{Könnecke et al., 2015}).
Data can also be exported in CSV format with the metadata exported in
JSON.

The stored data can be divided into distinct sections:

\begin{itemize}
\tightlist
\item
  Time-stamped raw data obtained during the execution of the measurement
  protocol.
\item
  Instrument settings.
\item
  Human-readable summary of the measurement protocol information (see
  \autoref{fig:h5_data}a).
\item
  Complete Python script that recorded the data (see
  \autoref{fig:h5_data}b) as well as information on the Python
  environment, i.e.~a list of used packages and versions.
\item
  User-defined metadata, e.g.~sample and user information.
\end{itemize}

\hypertarget{documentation}{%
\subsection{Documentation}\label{documentation}}

In-depth documentation and guides for installing, using and
troubleshooting can be found on the
\href{https://fau-lap.github.io/NOMAD-CAMELS/}{CAMELS documentation
webpage}
(\protect\hyperlink{ref-fuchsCAMELSConfigurableApplication}{Fuchs \&
Lehmeyer, n.d.}).

\hypertarget{acknowledgements}{%
\section{Acknowledgements}\label{acknowledgements}}

We thank Patrick Oppermann (Fritz-Haber-Institut der
Max-Planck-Gesellschaft) for valuable discussions.

NOMAD CAMELS is being developed within the NFDI consortium
\emph{FAIRmat} funded by the Deutsche Forschungsgemeinschaft ``DFG,
German Research Foundation'', project 460197019.

\hypertarget{references}{%
\section*{References}\label{references}}
\addcontentsline{toc}{section}{References}

\hypertarget{refs}{}
\begin{CSLReferences}{1}{0}
\leavevmode\vadjust pre{\hypertarget{ref-Allan2019}{}}%
Allan, D., Caswell, T., Campbell, S., \& Rakitin, M. (2019). {Bluesky's
Ahead: A Multi-Facility Collaboration for an a la Carte Software Project
for Data Acquisition and Management}. \emph{Synchrotron Radiat. News},
\emph{32}(3), 19--22.
\url{https://doi.org/10.1080/08940886.2019.1608121}

\leavevmode\vadjust pre{\hypertarget{ref-bluesky}{}}%
\emph{{Bluesky Project}}. (n.d.). Retrieved October 5, 2023, from
\url{https://blueskyproject.io/}

\leavevmode\vadjust pre{\hypertarget{ref-EPICS}{}}%
\emph{{EPICS - Experimental Physics and Industrial Control System}}.
(n.d.). Retrieved October 5, 2023, from
\url{https://epics-controls.org/}

\leavevmode\vadjust pre{\hypertarget{ref-FAIRmatNeXusProposal}{}}%
\emph{FAIRmat NeXus proposal}. (n.d.). Retrieved October 30, 2023, from
\url{https://fairmat-nfdi.github.io/nexus-fairmat-proposal/9636feecb79bb32b828b1a9804269573256d7696/fairmat-cover.html}

\leavevmode\vadjust pre{\hypertarget{ref-CAMELS_drivers}{}}%
{FAU-LAP/CAMELS{\_}drivers: device implementation for CAMELS}. (n.d.).
In \emph{GitHub repository}. GitHub. Retrieved September 28, 2023, from
\url{https://github.com/FAU-LAP/CAMELS_drivers}

\leavevmode\vadjust pre{\hypertarget{ref-DFG_position_paper}{}}%
\emph{{Förderung von Informationsinfrastrukturen für die Wissenschaft}}.
(2018).
\url{https://www.dfg.de/download/pdf/foerderung/programme/lis/positionspapier_informationsinfrastrukturen.pdf}

\leavevmode\vadjust pre{\hypertarget{ref-fuchsCAMELSConfigurableApplication}{}}%
Fuchs, A. D., \& Lehmeyer, J. A. F. (n.d.). {CAMELS} - {Configurable
Application} for {Measurements}, {Experiments} and {Laboratory Systems}
\textemdash{} {NOMAD-CAMELS} documentation. In \emph{NOMAD CAMELS -
Documentation}. Retrieved October 31, 2023, from
\url{https://fau-lap.github.io/NOMAD-CAMELS/}

\leavevmode\vadjust pre{\hypertarget{ref-Knott1994}{}}%
Knott, M., Gurd, D., Lewis, S., \& Thuot, M. (1994). {EPICS: A control
system software co-development success story}. \emph{Nucl. Inst. Methods
Phys. Res. A}, \emph{352}(1-2), 486--491.
\url{https://doi.org/10.1016/0168-9002(94)91577-6}

\leavevmode\vadjust pre{\hypertarget{ref-Konnecke2015}{}}%
Könnecke, M., Akeroyd, F. A., Bernstein, H. J., Brewster, A. S.,
Campbell, S. I., Clausen, B., Cottrell, S., Hoffmann, J. U., Jemian, P.
R., Männicke, D., Osborn, R., Peterson, P. F., Richter, T., Suzuki, J.,
Watts, B., Wintersberger, E., \& Wuttke, J. (2015). {The NeXus data
format}. \emph{J. Appl. Crystallogr.}, \emph{48}(1), 301--305.
\url{https://doi.org/10.1107/S1600576714027575}

\leavevmode\vadjust pre{\hypertarget{ref-pernstich2012}{}}%
Pernstich, K. P. (2012). Instrument {Control} ({iC}) \textendash{} {An
Open-Source Software} to {Automate Test Equipment}. \emph{Journal of
Research of the National Institute of Standards and Technology},
\emph{117}, 176--184. \url{https://doi.org/10.6028/jres.117.010}

\leavevmode\vadjust pre{\hypertarget{ref-PipDocumentationV23}{}}%
\emph{Pip documentation V23.3.1}. (n.d.). Retrieved October 30, 2023,
from \url{https://pip.pypa.io/}

\leavevmode\vadjust pre{\hypertarget{ref-PyMoDAQ}{}}%
\emph{PyMoDAQ}. (n.d.). Retrieved October 30, 2023, from
\url{https://pymodaq.cnrs.fr/}

\leavevmode\vadjust pre{\hypertarget{ref-PyVISA}{}}%
\emph{PyVISA}. (n.d.). Retrieved October 30, 2023, from
\url{https://pyvisa.readthedocs.io/}

\leavevmode\vadjust pre{\hypertarget{ref-scheidgenFAIRResearchData2023}{}}%
Scheidgen, M., Brückner, S., Brockhauser, S., Ghiringhelli, L. M.,
Dietrich, F., Mansour, A. E., Márquez, J. A., Albrecht, M., Weber, H.
B., Botti, S., Aeschlimann, M., \& Draxl, C. (2023). {FAIR Research Data
With NOMAD}: {FAIRmat}'s {Distributed}, {Schema-based Research-data
Infrastructure} to {Harmonize RDM} in {Materials Science}.
\emph{Proceedings of the Conference on Research Data Infrastructure},
\emph{1}. \url{https://doi.org/10.52825/cordi.v1i.376}

\leavevmode\vadjust pre{\hypertarget{ref-scheidgenNOMADDistributedWebbased2023}{}}%
Scheidgen, M., Himanen, L., Ladines, A. N., Sikter, D., Nakhaee, M.,
Fekete, Á., Chang, T., Golparvar, A., Márquez, J. A., Brockhauser, S.,
Brückner, S., Ghiringhelli, L. M., Dietrich, F., Lehmberg, D., Denell,
T., Albino, A., Näsström, H., Shabih, S., Dobener, F., \ldots{} Draxl,
C. (2023). {NOMAD}: {A} distributed web-based platform for managing
materials science research data. \emph{Journal of Open Source Software},
\emph{8}(90), 5388. \url{https://doi.org/10.21105/joss.05388}

\leavevmode\vadjust pre{\hypertarget{ref-SweepMe}{}}%
\emph{{SweepMe! - A multi-tool measurement software}}. (n.d.). Retrieved
September 28, 2023, from \url{https://sweep-me.net/}

\leavevmode\vadjust pre{\hypertarget{ref-Wilkinson2016}{}}%
Wilkinson, M. D., Dumontier, M., Aalbersberg, Ij. J., Appleton, G.,
Axton, M., Baak, A., Blomberg, N., Boiten, J. W., da Silva Santos, L.
B., Bourne, P. E., Bouwman, J., Brookes, A. J., Clark, T., Crosas, M.,
Dillo, I., Dumon, O., Edmunds, S., Evelo, C. T., Finkers, R., \ldots{}
Mons, B. (2016). {The FAIR Guiding Principles for scientific data
management and stewardship}. \emph{Sci. Data}, \emph{3}(1), 1--9.
\url{https://doi.org/10.1038/sdata.2016.18}

\end{CSLReferences}

\end{document}